\documentclass[aps,prx,floatfix,twocolumn,groupedaddress,superscriptaddress,nofootinbib,notitlepage]{revtex4-2}
\usepackage{times,graphics,graphicx,bm,bbm,bbold,amssymb,amsmath,amsfonts,dsfont,braket,hyperref}
\usepackage[utf8]{inputenc}
\usepackage[T1]{fontenc}
\hypersetup{colorlinks,linkcolor={blue},citecolor={blue},urlcolor={blue}}
\usepackage{float}
\usepackage{dblfloatfix}
\usepackage{amsmath,lineno}
\DeclareMathOperator*{\argmax}{arg\,max} 

\newcommand{\jtcom}[1]%
   {\begingroup{\color{blue}[JT: \textit{#1}]}\endgroup}
\begin{document}
\title{
State leakage during fast decay and control of a superconducting transmon qubit}
\author{ Aravind Plathanam  Babu}
\affiliation{QTF Center of Excellence, Department of Applied Physics, Aalto University, FI-00076 Aalto, Espoo, Finland}
\author{ Jani Tuorila }
\affiliation{QTF Center of Excellence, Department of Applied Physics, Aalto University, FI-00076 Aalto, Espoo, Finland}
\affiliation{IQM, Keilaranta 19, FI-02150 Espoo, Finland}
\author{ Tapio Ala-Nissila}
\affiliation{QTF Center of Excellence, Department of Applied Physics, Aalto University, FI-00076 Aalto, Espoo, Finland}
\affiliation{Interdisciplinary Centre for Mathematical Modelling and Department of Mathematical Sciences, Loughborough University, Loughborough, Leicestershire LE11 3TU, UK}
\date{June 9, 2020}

\begin{abstract}
Superconducting Josephson junction qubits constitute the main current technology for many applications, including scalable quantum computers and thermal devices. Theoretical modeling of such systems is usually done within the two-level approximation. However, accurate theoretical modeling requires taking into account the influence of the higher excited states without limiting \textcolor{black}{the system} to the two-level qubit subspace.  Here, we study the dynamics and control of a superconducting transmon using the numerically exact stochastic Liouville-von Neumann equation approach.  We focus on the role of state leakage from the ideal two-level subspace for bath induced decay and single-qubit gate operations.  We find significant short-time state leakage due to the strong coupling to the bath. We quantify the leakage errors in single-qubit gates and demonstrate their suppression with DRAG control for a five-level transmon in the presence of decoherence. Our results predict the limits of accuracy of the two-level approximation and possible intrinsic constraints in qubit dynamics and control for an experimentally relevant parameter set. 
\end{abstract}
\maketitle

\section{INTRODUCTION}

Recent developments in quantum devices are based on the high-fidelity control of two-level systems. Superconducting Josephson junction based qubits are the current leading choice for large scale quantum computing and devices such as quantum heat engines \cite{review, googlesycamore, quantumheatenginesReview}. 
Superconducting qubits are realized with different circuit designs, which are the charge, flux, and phase qubits \cite{koch}.
 These systems are ideally physical realizations of anharmonic multilevel systems in which the anharmonicity is caused by the inherent nonlinearity in the Josephson junction \cite{peter} and can be controlled by changing ratio of the Josephson to the charging energy.
Ideal qubit operation requires a strict two-level approximation which restricts the operational subspace into the two lowest energy eigenstates.

The population leakage from this low-energy subspace 
can induce significant error in the qubit dynamics and control. High-fidelity and error-free computing in particular requires a detailed understanding of the dynamics and control in the presence of these higher energy states for experimental circuits. Most importantly, coherent driving of qubits is the key requirement for the implementation of single-qubit quantum gates. Coherent driving of multilevel systems inherently induces excitations out of the qubit subspace \cite{steffen,multilevelrabiefffect} which, in addition to decoherence, limits the gate fidelities.  Optimization of leakage-free driving for single-qubit gates in the absence of decoherence 
has been previously investigated within the derivative removal adiabatic gates (DRAG) method which successfully reduces the leakage errors \textcolor{black}{\cite{fmotzoi,PRAchow,Mckay}}.

Further, applications of superconducting qubits are not limited to quantum computing. Superconducting qubits are also used as efficient quantum simulators and quantum heat engines, due to their high degree of controllability in preparation and readout \cite{quantumstimualtions_lamata,heatengin_newman}. A precise understanding of dynamics and control in the presence of higher energy states is important for proper operation of these devices.

Typically, studies in dynamics and control of superconducting qubits are restricted to weak coupling to the background or heat bath.  This \textcolor{black}{is } because most of the quantum computing devices require ultra-weak coupling between the qubit and the environment during gate operations. However, environmental engineering should not be restricted to weak coupling only. In particular, quantum heat engines and simulators may need relatively strong coupling to the bath \cite{review, strongcoupling_forn,  heatengin_newman, Non-MarkovianDynamicsofaQuantumHeatEngine}. 
Considerable attention has recently been focused on strong coupling regions in the context of fast qubit initialization with engineered and tunable environments \cite{NCqunatumcircuitrefrigerator,heatsink,jani,2017npjQI...3...27T}.
For a system interacting weakly with its environment the Markovian Lindblad  equation 
has proven to be remarkably successful in describing the dynamics of quantum devices \cite{breuer_OQS,gkl,lindblad}. However, for strongly interacting systems the Markovian weak-coupling approach can no longer be justified and more accurate methods must be employed. Among these methods, the stochastic Liouville-von Neumann (SLN) equation approach allows for 
a numerically exact solution of the reduced dynamics of the system with very few assumptions \cite{jurgen,Prljurgan}.

Here, we analyze the decay of a transmon qubit coupled to a bosonic bath and investigate how state leakage occurs during the decay using the exact SLN method, and the stochastic Liouville equation with dissipation (SLED) which is computationally efficient and equivalent to SLN in the limit of high cutoff for the environmental noise \cite{jurgen,Prljurgenandmak}. We show that the universal decoherence induces experimentally relevant short-time leakage during the decay. This indicates that when modeling transmon qubit decay dynamics, the two-level approximation may be inaccurate even at the zero temperature limit. We further study the influence of higher energy states in controlling the transmon, focusing in single qubit gate fidelities in the presence of a dissipative environment. 
We quantify leakage error in single-qubit quantum gates \textcolor{black}{and study the performance of DRAG control techniques} in the presence of decoherence for experimentally relevant parameters. Our studies thus quantify the influence of both higher energy levels and decoherence in the quantum gate operations and other applications relying on coherent qubit control protocols.

\section{RESULTS}
\subsection{System}
The effective Hamiltonian  of a superconducting charge qubit formed by a Josephson junction with Josephson energy $E_{\rm J}$ and charging energy $ E_{\rm C} $ can be defined as \cite{koch}
\begin{equation}
\hat H_{\rm q} =4E_{\rm C}(\hat n -n_{\rm g})^2-E_{\rm J} \cos\hat \phi,
\end{equation}
where $ n_{\rm g}$ is the effective offset charge number, and $ \hat n$ and  $\hat \phi$ are the net number of Cooper pairs transferred into the island and superconducting phase difference across the Josephson junction, respectively.
We approximate $\hat H_{\rm q}$ by truncating into the subspace spanned by its $N$ lowest energy eigenstates $|k\rangle$ as
\begin{equation}\label{eq:transHam}
    \hat H_{\rm q}=\hbar \sum_{k=0}^{N-1} \omega_k |k\rangle \langle k|,
\end{equation}
where $\omega_k $ \textcolor{black}{are the corresponding eigenfrequencies.}
In the following, we denote the lowest-transition angular frequency with $\omega_{01}=\omega_1-\omega_0$. In Ref. \onlinecite{jani} it has been shown that a truncation to $N=5$ lowest eigenstates is enough for accurate studies of single-excitation and low-temperature dynamics. In the so-called transmon regime of the charge qubit, $E_{\rm J}\gg E_{\rm C}$ and the lowest-energy eigenstates become independent on the offset charge number $n_{\rm g}$ which reduces the charge-noise sensitivity of the device. The typical transition frequency for a solid-state transmon is of the order of \textcolor{black}{4--}5 GHz  \textcolor{black}{with the absolute anharmonicity $\alpha=\omega_{12}-\omega_{01}$}  approximately around $-200$ MHz \cite{review,2017npjQI...3...27T}.

\subsection{ Bath induced decay and short-time decoherence} 

\textcolor{black}{
Transmon qubits are typically coupled to transmission line resonators for state read-out and control. The electromagnetic modes inside such resonator act as a bosonic dissipative environment for the transmon. We study such source of decoherence by considering}
the dynamics of a transmon
which is coupled to a bosonic bath at temperature $T$ with Hamiltonian $\hat H_{\rm B}=\Sigma_j\hbar \omega_j\hat b_j^\dagger \hat b_j$, where $ \hat b_j^\dagger$ and $ \hat b_j $ are the creation and annihilation operators of the bath oscillators. \textcolor{black}{Typically, the interaction between the transmon and the bath is conveniently modeled with a bilinear coupling. The interaction Hamiltonian can be written as} 
\begin{equation}
\hat H_{\rm I}(t)= \hbar \hat q\hat \zeta ,
\end{equation}
where $\hat q=\sum_{k,l}\langle k|\hat{n} |l\rangle |k\rangle \langle l|$ 
and $\hat \zeta=\Sigma_j g_j (\hat b_{j}^+ +\hat b_{j})$. 
\textcolor{black}{The  spectral characteristics of bath  can be modeled in terms of the spectral density
\begin{equation}
J(\omega)=2\pi \sum_{j} g_j^2\delta(\omega-\omega_j),
\end{equation}
where $g_j$ is the coupling angular frequency between the transmon and the bath oscillator $j$. In such superconducting circuits, the bosonic environments can be conveniently modeled with a resistor which has an ohmic spectral density } $J(\omega)=\kappa \omega/(1+\omega^2/\omega_{\rm c}^2)^2,$ where $\omega_{\rm c}$ is the cutoff frequency, and $\kappa$ is equal to the spontaneous emission rate of the Lindblad equation and thus determines the coupling strength.
\begin{figure}[h]
     \includegraphics[scale=0.5]{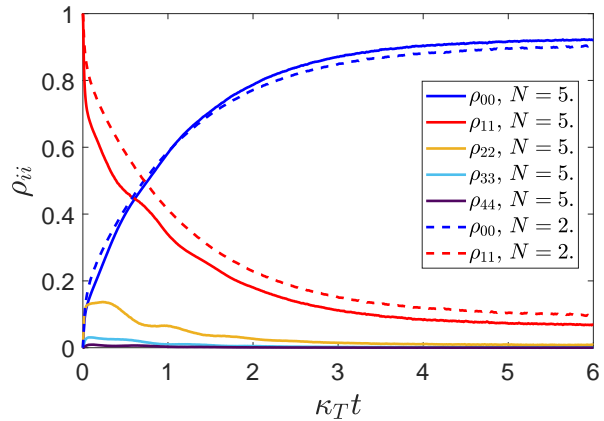}
    \caption{ Decay dynamics of the diagonal elements of the reduced density matrix of a transmon with $N$ states, starting from an occupied first excited state $\rho_{11}(t=0) = 1$. The data has been obtained using the SLED. We have used the parameters $E_{\rm J}/E_{\rm C}=100$, $\kappa/\omega_{01}= \textcolor{black}{0.2}$, $\beta \hbar \omega_{01}=5$, $\omega_{\rm c}=50\omega_{01} $,  and $\kappa_T=\kappa \coth(\beta \hbar \omega_{01}/2)$.} 
	\label{decay1}
\end{figure}

In the case of weak coupling, the interaction with the bath can be modeled accurately with the Lindblad equation. In particular, at low temperatures the leakage from the single-excitation subspace of the transmon is negligible and one can make a truncation to $N=2$ states in the Hamiltonian in Eq.~(\ref{eq:transHam}). 
However, for fast qubit initialization or a quantum heat engine a relatively strong coupling may be required, which warrants a detailed study of the expected corrections to the Lindblad results. In Ref. \onlinecite{jani}, the steady-state properties of a strongly coupled transmon beyond the two-level approximation have been studied in the context of fast qubit initialization.  Here, we focus on transient dynamics and the decay from the first excited state in particular, with an emphasis on leakage to states outside the qubit subspace.

Figure~\ref{decay1} shows 
the dynamics of the diagonal elements of the reduced density operator calculated with the SLED method for \textcolor{black}{ a transmon is initialized to the first excited state} \textcolor{black}{(see the Methods section for details on the SLN and SLED methods). We have used a relatively strong coupling $\kappa=0.2 \omega_{01}$ which can currently be realized with a tuneable environment \cite{heatsink, NCqunatumcircuitrefrigerator,hsu}. We emphasize that such strong coupling is not relevant for coherently operating transmon devices, but potentially relevant for fast qubit reset or quantum heat engines operating in the non-Markovian regime.  We choose $\beta \hbar \omega_{01} =5$, which corresponds to temperature around $38$ mK  for a transmon with frequency $\omega_{01}/2\pi=4$ GHz. This is close to typical experimental temperatures for transmon circuits. 
In Fig.~\ref{decay1}, we scale the time axis with the weak-coupling decay rate $\kappa_T=\kappa \coth(\beta \hbar \omega_{01}/2)$ to compare with the weak coupling decay.} \textcolor{black}{Fig.~\ref{decay1} demonstrates that }
at short times there is a significant population leakage to the higher excited states which is due to the universal decoherence described in Ref. \onlinecite{braun}.

We substantiate these numerical results and obtain a detailed description of the short-time decoherence by deriving an analytic solution in the early-time limit in which the free dynamics of the system can be neglected. \textcolor{black}{Details of the derivation can be found in Methods section.} We use the operator method described in Refs. \onlinecite{jani,braun} to obtain the diagonal elements as 
\begin{equation}\label{eq:de1}
    \langle0|\rho_S(t)|0\rangle=\frac{1}{6}[1-e^{-12f(t)\kappa/(4\pi \omega_{01})}];
\end{equation}
\begin{equation}\label{eq:de2}
    \langle1|\rho_S(t)|1\rangle=\frac{1}{2}[1+e^{-12f(t)\kappa/(4\pi \omega_{01})}];
\end{equation}
\begin{equation}\label{eq:de3}
    \langle2|\rho_S(t)|2\rangle=\frac{1}{3}[1-e^{-12f(t)\kappa/(4\pi \omega_{01})}],
\end{equation}
where
\begin{equation}
   f(t)=\frac{2\omega_{01}}{\kappa}\int_0^{\infty}d\omega \frac{J(\omega)}{\omega^2} \coth(\hbar\beta\omega/2)[1-\cos(\omega t)].
\end{equation}
%
%
We have set $N=3$ here 
\textcolor{black}{which makes the analytic calculations feasible and reveals the main factors contributing to leakage. We have} assumed that the transmon is initially in the first excited state, similar to the numerical data shown 
in Fig. \ref{decay1}. 
\begin{figure}[H]
     \includegraphics[scale=0.5]{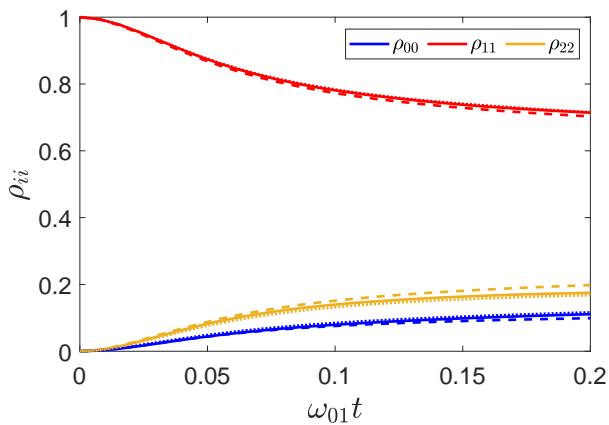}
    \caption{Short-time dynamics of the diagonal elements of the density matrix for decay from the first excited state calculated with SLN (solid lines), SLED (dotted lines) and our analytic solution in Eqs.~(5)-(7) (dashed lines). We have used $E_{\rm J}/E_{\rm C}=100$, $\kappa/\omega_{01} =\textcolor{black}{0.2}$, $\beta \hbar \omega_{01}=5 $, $\omega_{\rm c}=50\omega_{01} $ and $N=3$.} 
	\label{shorttime-decay}
\end{figure}
Figure \ref{shorttime-decay} shows a detailed comparison between our analytic solution for the short-time decay and the corresponding numerical data obtained with the SLN and SLED methods. The excellent agreement validates the conclusion that the short-time leakage in the transmon is due to the universal short-time decoherence. We note that Ref. \onlinecite{jani} also reports analytic results for short-time dynamics but only in the two-level approximation for the transmon.\\

\begin{figure}[ht]
     \includegraphics[scale=0.5]{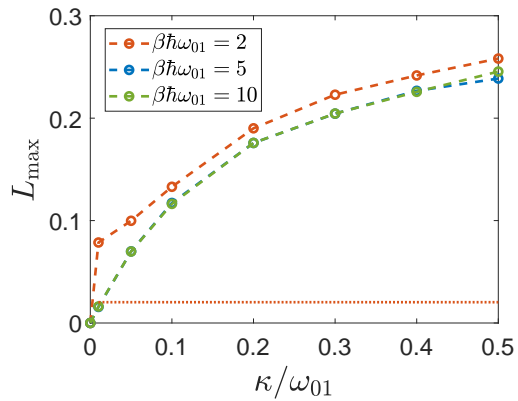}
    \caption{ Maximum state leakage for the decay from the first excited state. \textcolor{black}{Note that blue and green lines almost overlap each other. } Dotted lines indicate the steady-state thermal leakage calculated using the Boltzmann distribution for $\beta \hbar \omega_{01}=2$.
    Here $E_{\rm J}/E_{\rm C}=100$, $\omega_{\rm c}=50\omega_{01} $ and $N=5$.}
	\label{decay-lmax}
\end{figure}
\begin{figure}[H]
     \includegraphics[scale=0.5]{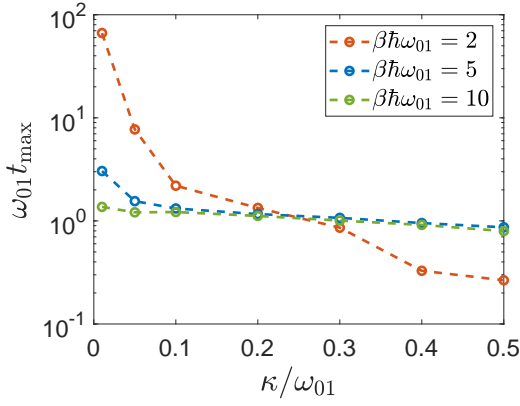}
    \caption{\textcolor{black}{  The time at which maximum state leakage occurs for  different values of $\kappa/\omega_{01}$}. The system is  initially in the first excited state. 
    Here $E_{\rm J}/E_{\rm C}=100$, $\omega_{\rm c}=50\omega_{01} $ and $N=5$.} 
	\label{decay-tmax}
\end{figure}

The population leakage demonstrated in Fig.~\ref{decay1} can be quantified in terms of state leakage $L$ of the density matrix as \cite{wood}
\begin{equation}
    L(\rho)=1-{\rm Tr}[P_1\rho],
\end{equation}
where $P_1=|0\rangle \langle 0|+|1\rangle \langle 1|$ is the projector onto the computational (ideal) subspace and $\rho$ is the density matrix of the system. 
\textcolor{black}{The maximum of state leakage over time $L_{\rm max} $ for a 5-level transmon is shown in Fig. \ref{decay-lmax}. The data reveal that the state leakage is a monotonically increasing function of the coupling constant $\kappa$. As expected, the maximum leakage at high temperatures significantly increases due to thermal excitations. It is worth noting that the maximum state leakage is larger than the state leakage in thermal equilibrium, 
and we emphasize that at finite temperatures the Lindblad equation predicts that the leakage increases monotonically towards the steady-state value given by the Boltzmann distribution.} \textcolor{black}{In order to understand how the maximum leakage depends on temperature, we define the time at which $L_{\rm max}$ occurs as
\begin{equation}
    t_{\rm max }= \argmax_{t} L (\rho).
\end{equation}
In Fig. \ref{decay-tmax}, we show $t_{\rm max }$ for different values of $\kappa$ and $\beta$.}
 In the case of strong coupling ($\kappa>0.1 \omega$) short-time decoherence dominates and maximum leakage occurs at early times. \textcolor{black}{However,}
 for a relatively weak coupling ($\kappa<0.1 \omega$) the thermal excitations start to dominate and the time at which the maximum leakage occurs is approximately $100/ \omega_{01}$. Thus, we expect that the short-time leakage could be studied experimentally for moderate temperatures and bath coupling strengths using fast measurement techniques.

\textcolor{black}{These results imply that higher-order levels are essential for accurately modeling the transmon dynamics even at low temperatures and, thus, 
should be taken carefully into account when modeling devices that use strong and possibly controllable couplings to the environment.}

\subsection{Control with classical driving}

Coherent driving of qubits is crucial for the realization of quantum logical operations.  To this end,
we consider the case of an isolated transmon driven unitarily by an external classical field as
\begin{equation}\label{eq:drivenHam}
        \hat H_{\rm S}=\hbar \sum_{k}^{N-1}\omega_k |k\rangle \langle k|+\hbar \Omega \hat q  \cos(\omega_{d}t),
\end{equation}
where $\Omega$ 
determines the amplitude of the cosine pulse and $\omega_d$ is the is the driving frequency.
\textcolor{black}{In order to reduce the complexity of the analytic calculations, } we consider here the \textcolor{black}{idealized case of a} 3-level transmon ($N=3$) \textcolor{black}{in the absence of dissipation. In the sections below, we use $N=5$ in our numerical simulations which quantify leakage errors in single-qubit gate operations.}

We transform the Hamiltonian into the rotating frame \textcolor{black}{using} 
the unitary operator
$ \hat U=e^{i\omega_{d}t \hat a^{\dagger} \hat a}$, where \textcolor{black}{$\hat a= |0\rangle \langle 1|+\epsilon |1\rangle \langle 2|$ and $\epsilon=\langle 1|\hat n |2\rangle$ and $\epsilon \approx \sqrt{2} $.} 
\textcolor{black}{
\begin{equation}
\begin{split}
 \Tilde{H}_{\rm S}&=\hat U \hat H_{\rm S} \hat U^{\dagger}+i\hbar \hat U \dot{\hat U}^{\dagger},\\& =\hbar \begin{bmatrix}
0     & \Omega f(t)& 0 \\
\Omega f(-t)& \omega_{\rm 01}- \omega_{\rm d}& \sqrt{2} \Omega
f(-t)\\
0& \sqrt{2} \Omega f(t)& \omega_{\rm 02}-2 \omega_{\rm d}
\end{bmatrix},
\end{split}
\end{equation}
where $f(t)=e^{-i\omega_{\rm d}t} \cos(\omega_{\rm d}t)$.}
\textcolor{black}{In order to simplify further we make the rotating wave approximation by neglecting the fast oscillating terms and assume resonance condition $\omega_{\rm 01}=\omega_{\rm d}$. We obtain
\begin{equation}\label{eq:RWAHam}
\Tilde{H}_{\rm S}=\hbar \begin{bmatrix}
0     & \Omega/2& 0 \\
\Omega/2& 0& \sqrt{2} \Omega/2\\
0&\sqrt{2} \Omega/2& \omega_{\rm 02}- 2\omega_{\rm d}
\end{bmatrix}.
\end{equation}}
\begin{figure}[ht]
 \centering
	\includegraphics[scale=0.5]{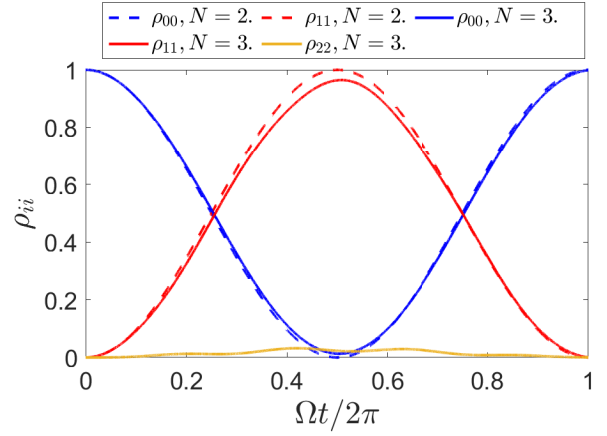}
     \label{rabi0}
\caption{Dynamics of the diagonal elements of the density operator of a driven transmon in the rotating frame. The transmon is initially in the ground state.  The parameters here are $ \Omega =0.01\omega_{01}$, and $\omega_{d}=\omega_{\rm 01}$.}
\label{rabi-plot}
\end{figure}

\textcolor{black}{In the case studied here, the system is isolated and the dynamics of the diagonal elements of the density operator can be calculated by solving the von Neumann equation. Figure \ref{rabi-plot} shows the dynamics of the diagonal elements of the density operator calculated by solving the von Neumann equation for the Hamiltonian in  Eq.~(\ref{eq:RWAHam})}  with $ \Omega =0.01\omega_{01}$ and $\omega_{\rm d}=\omega_{01}$.
The data reveal that the presence of the second excited state induces state leakage during the Rabi oscillations, which can lead to significant leakage error in single-qubit gate operations.  In the following section, we 
discuss this in detail for the case of a single-qubit  NOT-gate.  \textcolor{black}{Again, the data imply that the two-level approximation is not accurate for modeling the dynamics of a driven transmon.} 
\begin{figure}[ht]
 \includegraphics[scale=0.5]{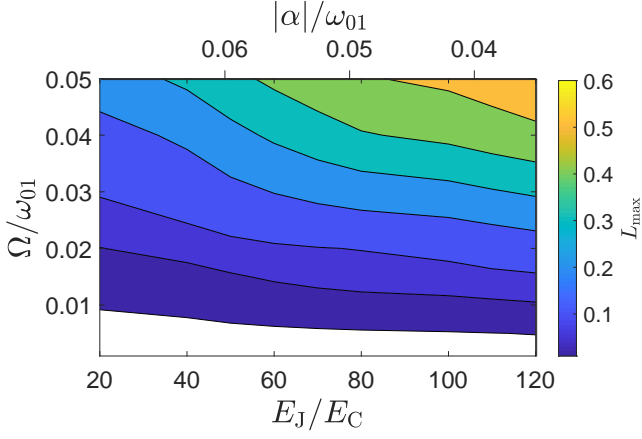}
    \caption{Maximum state leakage for three-level transmon as a function of the relative drive coupling $ \Omega /\omega_{01}$ and energy ratio $E_{\rm J}/E_{\rm C}$ \textcolor{black}{or $ |\alpha|/\omega_{01}$}, corresponding to the $N=3$ transmon in Fig. V. \textcolor{black}{Where $\alpha=\omega_{12}-\omega_{01}$ and  the white region corresponds to $L_{\rm max}<0.01$.}}
   \label{rabi4}
\end{figure}
In Fig.~\ref{rabi4}, we show 
the relationship between the \textcolor{black}{relative anharmonicity  $\alpha/\omega_{01}$} (which is inversely proportional to $\sqrt{E_{\rm J}/E_{\rm C}}$ in the transmon limit), the amplitude $\Omega$ of driving, and the maximum state leakage $L_{\rm max}$. \textcolor{black}{Note that here we calculated maximum state leakage during one Rabi cycle and assumed that the transmon is initially in the ground state.} 
Clearly, decreasing anharmonicity enhances state leakage as the \textcolor{black}{transitions between adjacent excited states become closer to resonance with the drive.} 
The data illustrate the intrinsic limits for the values of $\Omega$ and $E_{\rm J}/E_{\rm C}$ to reduce the maximum leakage to $L_{\rm max}<0.01$ (white region in the plot). \textcolor{black}{Driving a transmon with a small amplitude obviously reduces leakage errors. However, we need to consider the effect of the dissipative environment. Its influence is expected to be relatively weak in the weak-coupling limit, but it still could have notable effects in devices that require high accuracy.}

\subsection{ Control with single qubit gates }

\begin{figure}[h!]
     \centering
      \includegraphics[scale=0.5]{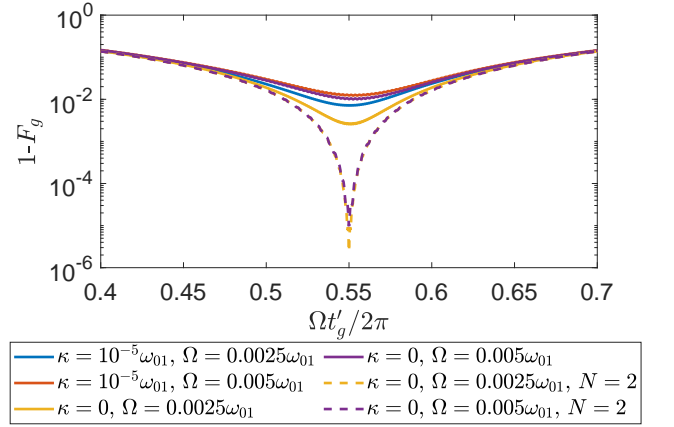}
        \caption{The infidelity
        of a simple NOT-gate with different drive amplitudes and bath coupling strengths calculated using the Redfield equation for $N=5$ (solid lines) and $N=2$ (dashed lines). Here $t'_{\rm g}=t_{\rm g}+2 t_{\rm r} $ and we have used  $\beta\hbar \omega_{01}=10$, $E_{\rm J}/E_{\rm C}=100$  and $t_{\rm r}=t_{\rm g}/20$.} 
        \label{gateop1}
    \end{figure}

Finally, we study the influence of dissipation to the operation of optimized single-qubit gates. 
Superconducting single qubit gates are constructed with various type of controls. A general gate operation for a $N$-level transmon can be written as 
\begin{equation}
    H=\hbar \sum_{k}^{N-1}\omega_k |k\rangle \langle k|+\hbar  \hat q \varepsilon(t),
\end{equation}
with
\begin{equation}
   \varepsilon(t)=\begin{cases}
    \varepsilon_x(t) \cos(\omega_d t)+\varepsilon_y(t) \sin(\omega_d t), & \text{for $0<t<t_g$};\\
    0, & \text{otherwise},
  \end{cases}
\end{equation}
where $\varepsilon_x(t),\varepsilon_y(t)$ are \textcolor{black}{mutually} independent quadrature controls, and $t_g$ is the gate time. For a simple NOT-gate, we choose \textcolor{black}{
\begin{equation}
   \varepsilon_x(t)=\begin{cases}
  \Omega R(t), &\text{for  $0 <t \leq t_{\rm r}$ };\\
    \Omega, & \text{for $t_{\rm r} \leq t<t_{\rm g}+t_{\rm r}$};\\
   \Omega R(t-t_{\rm g}+2t_{\rm r}), &\text{for  $t_{\rm g}+t_{\rm r} <t \leq t_{\rm g}+2t_{\rm r}$ },\\
  \end{cases}
\end{equation}
and $\varepsilon_y(t)=0$. Here, $R(t)=[\cos(\cos(\pi t/2t_{\rm r}) )- \cos(1)]/[1 - \cos(1)]$ is the ramping function, and $t_r$ is the ramp time to and from the constant value of $\Omega$. As a consequence, the protocol mimics a typical experimental situation in which the change of parameters has to have a finite rate.}\\
The average fidelity of single qubit gate operations can then be defined as \textcolor{black}{\cite{fmotzoi}}
\begin{equation}
\begin{split}
     F_{\rm g}=\frac{1}{6}\sum_{j=\{\sigma_z^{\pm},\sigma_x^{\pm},\sigma_y^{\pm}\}}{\rm Tr}[U_{\rm ideal}\rho(0)U^{\dagger}_{\rm ideal}\rho(t_{\rm g})],
\end{split}
\end{equation}
where $U_{\rm ideal} $ is the unitary operation for the corresponding gate operation in an ideal basis and \textcolor{black}{$\sigma_{i=x,y,z}^{\pm}$ are the eigenstates of the corresponding operators $\sigma_i$}. We estimate the gate error in terms of infidelity \textcolor{black}{in the presence of a} 
coupling to a bosonic bath at temperature  $T$. We set the coupling \textcolor{black}{frequencies with the bath oscillators to very small values, resulting to ultra-weak decay rates $\kappa$ for the transmon.} \textcolor{black}{
Instead of Linblad equation, we model the dissipative dynamics with the Redfield equation in order to avoid errors arising from the secular approximation (see Methods). We also note that earlier work has demonstrated that in the weak-coupling limit both SLED and Redfield results agree for the gate operations \cite{jani}}.\\
In Fig. \ref{gateop1}, we show the average infidelity ($1-F_{\rm g}$) calculated with the Redfield equation for the case of an isolated and weakly coupled transmon with simple  NOT-gate driving. 
We emphasize that the optimal gate time $t_{\rm g}=\pi/\Omega $ with ramp time $t_{\rm r}=t_{\rm g}/20$ is independent of the coupling strength $\kappa$, but the corresponding value of infidelity depends on $\kappa$ and $\Omega$. The difference between the optimal fidelity for $N=5$ (solid lines) and  $N=2$ (dashed lines) illustrates the error due to the presence of the higher energy states. \textcolor{black}{The data further demonstrate the variation in the optimal fidelity due to the coupling with the environment.  Note that dissipation leads to errors in single-qubit gate operation even in the weak coupling and low-temperature limit.}

It has been 
shown that the derivative removal by the adiabatic gate (DRAG) significantly suppresses the leakage errors in single-qubit operations \cite{fmotzoi}. In DRAG control one uses a Gaussian pulse for $\varepsilon_x(t)$ given by \textcolor{black}{
\begin{equation}
    \varepsilon_x(t)=  \Omega A [e^{-(t-t_g/2)^2/(2t_g^2)} -e^{-(t_g/2)^2/(2t_g^2)}],
\end{equation}}
with $ \varepsilon_y(t)=- \dot{\varepsilon}_x(t)/\alpha$. The amplitude $A$ determines the desired rotation (here we use $\pi$ for the NOT-gate) and $\alpha$ is the anharmonicity. 
       \begin{figure}[ht]
     \centering
      \includegraphics[scale=0.5]{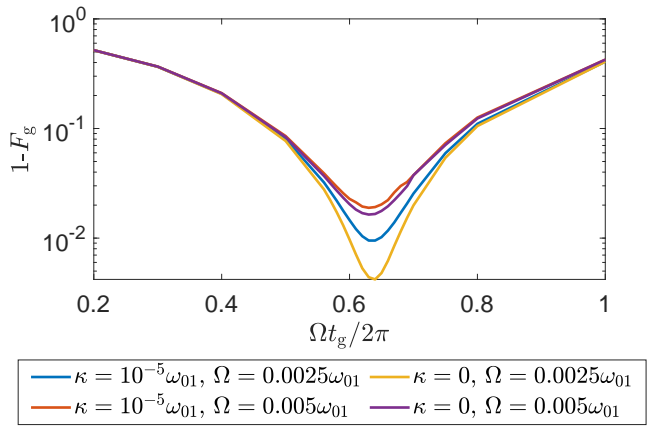}
        \caption{ The infidelity 
        of the DRAG NOT-gate calculated using Redfield equation. We have used $N=5$, $E_{\rm J}/E_{\rm C}=100$ and $\beta\hbar \omega_{01}=10$.} 
        \label{gatedrag}
    \end{figure}
Figure \ref{gatedrag} shows the average infidelities $1-F_{\rm g}$ calculated using the Redfield equation for an isolated and weakly coupled transmon NOT-gate with DRAG  driving. Naturally, the infidelity grows with gate speed and dissipation which cause leakage and decoherence errors, respectively. Note that the optimal gate time is changed from  $t_{\rm g}=\pi/\Omega $ due to the Gaussian pulse shape, but is still independent of the coupling strength $\kappa$. \textcolor{black}{The optimal values of the infidelity for simple and DRAG NOT-gates depend on $\kappa$ and $\Omega$. These results quantitatively confirm the importance of considering higher states and dissipative effects while modeling the single-qubit gate operation for a transmon qubit. We emphasize that the infidelities around $10^{-2}$ are not optimal for typical transmon gates but more reliable optimization of pulse shapes and preventing dissipative effects are required for realistic quantum computing applications.}\\

We improve our estimates of the leakage error by quantifying  
it in terms of average state leakage 
\begin{equation}
\begin{split}
     \Bar{L}_{\rm g}=\frac{1}{6}\sum_{j=\{\sigma_z^{\pm},\sigma_x^{\pm},\sigma_y^{\pm}\}} L[\rho^j(t_{\rm g})],
\end{split}
\end{equation}
where $L[\rho^j(t_{\rm g})]$ is the state leakage corresponding to the gate operation on the initial state $j$. 
\begin{figure}[ht]
     \centering
      \includegraphics[scale=0.5]{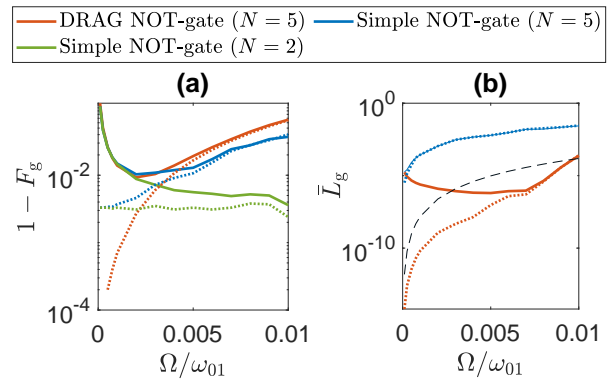}
        \caption{ (a) The average infidelity of 
        the simple and DRAG NOT-gates. (b) Average state leakage for the simple and DRAG NOT-gates. \textcolor{black}{The black dashed lines represents the curve for $\Omega^4/\alpha^3$}. We have used $E_J/E_C=100$, $\beta\hbar \omega_{01}=10$, $\kappa=10^{-5}\omega_{01}$(solid lines) and $\kappa=0$ (dotted lines) .}
        \label{gatedrag_simple}
    \end{figure}
In Fig. \ref{gatedrag_simple}, we study the effect of dissipation to the average infidelity and average state leakage of simple and DRAG-controlled NOT gates.
The presence of a weakly dissipative environment significantly reduces the gate fidelities of both gates for relatively small values of driving amplitude $\Omega$ and decoherence effects are negligible for the case of relatively fast gate operations (large values of  $\Omega $). \textcolor{black}{The errors due to the dissipative environment can be reduced by fast driving. However, fast driving can increase leakage errors. Ideally, the combination of a moderate driving amplitude and isolating the system from the bath(s) could optimize fidelity. The point where the infidelity has a minimum corresponds to the optimal driving amplitude for the parameters used in our work.} The average state leakage in the DRAG NOT-gate increases due to the dissipative environment and leads to a smaller gate fidelity. The DRAG NOT-gate is robust against the state leakage error and eliminates leakage to order $\Omega^4/\alpha^3$. \textcolor{black}{The state leakage is always less than or of the same order as $\Omega^4/\alpha^3$ (black dashed lines) for the isolated case.} This agrees with the result reported in Ref. \onlinecite{fmotzoi} for the case of an isolated qubit with $N=3$.  Also, our data clearly show that the DRAG pulse transfers the leakage errors to bit- and phase-flip errors, which can be seen as a multiple orders-of-magnitude difference between the simulated simple and DRAG average state leakages in Fig.~\ref{gatedrag_simple}(b). \textcolor{black}{Such errors can be corrected with quantum-error correcting codes.}
Note that the gate fidelity for the DRAG NOT-gate is smaller than that of the simple NOT-gate for large $\Omega $ but the pulse shape for DRAG can be optimized to obtain higher fidelity \cite{globaloptimization, chow}. Our data show that for non-optimized pulses the bit and phase-flip errors can be of the same order of magnitude as that caused by decoherence.  We emphasize that in our simulations we have taken into account both the dissipation and presence of higher energy states up to $N=5$ for experimentally relevant parameters. 

\section{DISCUSSION}

In summary, we have presented a detailed study of the state leakage of a transmon under strong dissipation and fast coherent control. We have studied the short-time decay dynamics 
of a 5-level transmon qubit coupled to a bosonic bath using analytic and numerically exact methods. Our results demonstrate that the universal decoherence induces significant short-time leakage during the decay. At experimentally relevant low temperatures, the leakage is directly proportional to the coupling to the bath. 
We found that the two-level approximation
of a transmon qubit does not result into accurate dynamics, even in the case of low temperature and small excitation number, 
especially in the case of systems which are strongly coupled to the environment. We expect this to be important if strong coupling is required for environmental control e.g. in qubit reset or in the case of quantum heat engines.

Furthermore, we have
quantified the dependence between state leakage, anharmonicity and drive amplitude for a resonantly driven qubit, and predicted parameter values relevant for minimal leakage. 
Finally, we have illustrated leakage errors in single-qubit quantum gates for a five-level transmon in the presence of decoherence, and suppression of leakage through DRAG control techniques for experimentally relevant parameters.  We have also predicted fidelity variations in single-qubit gates for a five-level transmon in the presence of decoherence for simple and DRAG-controlled NOT-gates. 
\section{METHODS}
   \subsection{Models for dissipation}
   
    In the main text, we have studied 
    open quantum system dynamics of a transmon bilineraly coupled to a bosonic bath. 
    We have modeled the setup with the Hamiltonian 
\begin{equation}
\hat H_{\rm SB}=\hat H_{\rm S}+\hat H_{\rm B}+\hat H_{\rm I},
\end{equation} 
where $\hat H_{\rm S}$ and $\hat H_{\rm B}$ are the Hamiltonians of the system and the bath, respectively.  
The interaction Hamiltonian is defined as 
\begin{equation}
\hat H_{\rm I}= \hbar \hat q\hat \zeta.
\end{equation}
The dynamics of the system is determined by its reduced density operator $\hat \rho_{\rm S} = \textrm{Tr}_{\rm B}~\hat \rho$, where $\hat \rho$ is the joint density operator of the system and the bath, the time evolution of which is determined by the von Neumann equation.

\subsubsection{Redfield and Lindblad equations}

If the coupling to the bath is weak, one can apply the conventional perturbative approach which includes 
Born and Markov approximations. As a result, we obtain the Redfield equation \cite{breuer_OQS} 
	\begin{equation}
	\begin{split}
	\frac{ d\hat\rho_{\rm S}(t)}{dt}=-\frac{i}{\hbar}[\hat H_{\rm S},\hat \rho_{\rm S}(t)]&+\frac{1}{2}\sum_{nmkl}  \Big \lbrace [S(\omega_{nm})+S(-\omega_{kl})]\\&\Pi_{nm}\hat\rho_{\rm S}\Pi_{kl}-S(\omega_{kl})\Pi_{nm}\Pi_{kl}\hat\rho_{\rm S}\\&-S(-\omega_{nm})\hat\rho_{\rm S}\Pi_{nm}\Pi_{kl}\Big\rbrace,
	\label{re12}
	\end{split}
	\end{equation}
	where $\omega_{nm} = \omega_m-\omega_n$ with $\omega_n$ being the eigenfrequencies of the system, $\Pi_{nm}=q_{nm}\vert n\rangle\langle m\vert$, \textcolor{black}{and $q_{nm}=\langle n| \hat q |m \rangle $ .} 
	We have replaced the correlation function by the inverse Fourier transform of spectrum of environmental fluctuations [$S(\omega)$]
	\begin{equation}
\langle \hat \zeta(t)\zeta(0)\rangle=\frac{1}{2\pi}\int_{-\infty}^{+\infty} d\omega e^{-i\omega t}  S(\pm \omega).
	\label{re2}
	\end{equation}
	For \textcolor{black}{an odd spectral density [$J(-\omega) = -J(\omega)$], such as spectral density of an ohmic bath}, we obtain
	\begin{equation}
	   S(\omega)= \frac{2J(\omega)}{1-e^{-\hbar\beta\omega }}.
	\end{equation}
\textcolor{black}{The Redfield  master equation can be reduced into Lindblad form by applying secular approximation by including only terms for which $\omega_{nm}+\omega_{kl}=0$ (equivalent to the rotating wave approximation ).
	 Consequently, one obtains
	\begin{align}
	\frac{ d\hat\rho_{\rm S}(t)}{dt} =&\frac{-i}{\hbar}[\hat H_{\rm S},\rho_{\rm S}]
	+\frac{1}{2}\sum_{\omega_{nm}>0}S(-\omega_{nm})\nonumber\\& \big[2\Pi_{nm}\hat\rho_{\rm S}\Pi_{nm}^\dagger-\lbrace \Pi_{nm}^\dagger\Pi_{nm},\hat\rho_{\rm S}\rbrace\big]\nonumber\\& +\frac{1}{2}\sum_{\omega_{nm}>0}S(\omega_{nm})\nonumber\\&\big[2\Pi_{nm}^\dagger\hat\rho_{\rm S}\Pi_{nm}-\lbrace \Pi_{nm}\Pi_{nm}^\dagger,\hat\rho_{S}\rbrace\big]\nonumber\\&+  \frac{1}{2}\sum_{n}S(0)\nonumber\\&[2\Pi_{nn}\hat\rho_{S}\Pi_{nn}-\lbrace \Pi_{nn}\Pi_{nn},\hat\rho_{\rm S}\rbrace],  
	\label{re14}  
	\end{align} 
	where $ S(0)= \lim_{\omega \to 0} S(\omega)=\kappa /\hbar \beta \omega_{01}$.}

\subsubsection{SLN and SLED equations}

If the coupling to the bath is strong, the above perturbative approximation becomes inaccurate. Formally, the dynamics can be solved in the path-integral formalism, but in practise the solution becomes untractable. In the case of a bilinearly-coupled Gaussian bath, one can reorganize the path-integral representation into the form of a stochastic Liouville--von Neumann equation (SLN) which can be solved efficiently\textcolor{black}{, at least in low-dimensional Hilbert spaces}. %

The SLN equation for the reduced density operator of the system can be written into the form~\cite{jurgen,Prljurgan}
    \begin{equation}
        i \hbar \frac{d \hat \rho_{\rm S}(t)}{dt}=[\hat H_{\rm S},\hat \rho_{\rm S}(t)]-\xi(t) [\hat q,\hat \rho_{\rm S}(t)]-\frac{\hbar}{2} \nu(t)\{\hat q ,\hat \rho_{\rm S}(t)\},
        \label{SLN}
    \end{equation}
    where $\xi$ and $\nu$ are complex noise terms encoding the correlations between the system and the bath. 
    These complex noise terms have to fulfill the correlation functions
    \begin{equation}
        \begin{split}
            &\langle \xi(t)\xi(t') \rangle={\rm Re}[\langle\zeta(t)\zeta(t')\rangle];\\&
            \langle \xi(t)\nu(t') \rangle=i \Theta(t-t') {\rm Im}[\langle\zeta(t)\zeta(t')\rangle];\\&
            \langle \nu(t)\nu(t') \rangle=0,
        \end{split}
    \end{equation}
    where $\Theta(t-t')$ is the Heaviside step function and the bath correlation function
    \begin{equation}
    \begin{split}
        \langle \xi(t) \xi(t') \rangle =\int_0^{\infty}& \frac{d\omega}{2 \pi}J(\omega)\{\coth[\hbar \beta \omega /2] \\&\times \cos[\omega(t-t')]-i\sin[\omega(t-t')]\}.
    \end{split}
    \end{equation}
    In our calculations, we use the ohmic spectral density with a Drude cutoff as  $J(\omega)=\kappa \omega/(1+\omega^2/\omega_{\rm c}^2)^2,$ where $\omega_{\rm c}$ is the cutoff frequency. 
    
    If the cutoff frequency  $\omega_{\rm c}$ is much larger than qubit frequency $\omega_{01}$, the SLN equation can be written
into the form of SLED as \textcolor{black}{\cite{Prljurgan,jani,Prljurgenandmak}}
 \begin{equation}
 \begin{split}
     \frac{d \hat \rho_{\rm S}(t)}{dt}=&-\frac{i}{\hbar}[\hat H_{\rm S},\hat \rho_{\rm S}(t)]+i \xi(t) [\hat q,\hat \rho_{\rm S}(t)]\\&-\frac{\kappa}{2\hbar \beta\ \omega_{01}} \big[\hat q ,[\hat q, \rho_{\rm S}(t)]\big]-\frac{i\kappa}{4}\big[\hat q ,[\hat p, \rho_{\rm S}(t)]\big],
        \label{SLED}
 \end{split}
    \end{equation}
    where $\hat p$ is the canonical conjugate of $\hat q $. We emphasize that the above SLN and SLED equations treat the interaction with the bath in a formally exact manner, but with the expense that they are stochastic. Therefore, the dynamics of the system density operator has to be solved for several realizations of the correlated noise terms.
    Moreover, a time-trajectory of the density operator given by an individual noise realization is unphysical, but physical results can be obtained by averaging over the realizations. 
    In our calculations, we typically average over $10^5$ realizations of the noise.

\subsection{Analytic solution for short-time dynamics}

Here, we derive an analytic expression for short time decoherence of a three-level transmon shown \textcolor{black}{in Eqs.~(\ref{eq:de1})--(\ref{eq:de3})}. In the early time limit, one can neglect the intrinsic dynamics of the system ($\hat H_{\rm S}$) and, as a result, one can write the elements of the reduced density matrix in the eigenbasis of the operator $\hat q $ as  \cite{jani}
\begin{equation}
\begin{split}
    \rho_{nm}(t)=&\exp([-(n-m)^2f(t)\\&
    +i(n^2-m^2)\phi(t)]\kappa/4 \pi \omega_{\rm q}) \rho_{nm}(0),
\end{split}
\end{equation}
where 
$$\rho_{nm}(t) =\langle n|\rho_{\rm S}(t)|m\rangle;$$
$$
   f(t)=\frac{2\omega_q}{\kappa}\int_0^{\infty}d\omega \frac{J(\omega)}{\omega^2} \coth(\hbar\beta\omega/2)[1-\cos(\omega t)];
$$
$$
   \phi(t)=\frac{2\omega_q}{\kappa}\int_0^{\infty}d\omega \frac{J(\omega)}{\omega^2} )[\omega t-\sin(\omega t)],
$$
and the spectral density 
$$
    J(\omega)=\frac{\kappa (\omega/\omega_q)}{(1+\omega^2/\omega_c^2)^2}.
$$
For a three-level transmon system, i.e. a qutrit, the operator  $\hat q $ can be approximated as
\begin{equation}
     \hat q =\begin{bmatrix} 
0 & 1& 0  \\
1 &0&\sqrt 2\\
0 &\sqrt 2&0 \\
\end{bmatrix},
\end{equation}
the eigenvectors of which are   
$\big\{ |q_{\pm}\rangle=(1/\sqrt{6},\pm 1/\sqrt{2},1/\sqrt{3} )^{\rm T},\: |q_{0}\rangle=(-\sqrt{2/3},0,1/\sqrt{3})^{\rm T}\big\}$. 
We 
express these eigenvectors in  the eigenbasis of the system Hamiltonian as
\begin{equation}
\begin{split}
   & |\hat q_0\rangle=-\sqrt{\frac{2}{3}}|0\rangle+\frac{1}{\sqrt{3}}|2\rangle;\\&
    |\hat q_\pm\rangle=\frac{1}{\sqrt{6}}|0\rangle| \pm\sqrt{\frac{1}{2}}|1\rangle+\frac{1}{\sqrt{3}}|2\rangle.
\end{split}
\end{equation}
Thus,
\begin{equation}
\begin{split}
    &|0\rangle=\frac{|\hat q_+\rangle+|\hat q_-\rangle-2|\hat q_0\rangle}{\sqrt 6};\:\:\:\:\: |1\rangle=\frac{|\hat q_+\rangle-|\hat q_-\rangle}{\sqrt 2};\\& |2\rangle=\frac{|\hat q_+\rangle+|\hat q_-\rangle+|\hat q_0\rangle}{\sqrt 3}.
\end{split}
\end{equation}
With these simplifications, we calculate the elements of the reduced density matrix in the eigenbasis of the system Hamiltonian as
\begin{equation}
\begin{split}
    \rho_{00}(t)=&\frac{1}{6}\Big[2\rho_{q_0q_0}(0)+ \rho_{q_+q_+}(0)+\rho_{q_-q_-}(0)\\&+e^{-12f(t)\kappa/(4\pi \omega_{\rm q})}[ \rho_{q_+q_-}(0) +\rho_{q_+q_-}(0)]\\&-2e^{-3f(t)\kappa/(4\pi \omega_{\rm q})}\big \{e^{3i\phi(t)\kappa/(4\pi \omega_{\rm q})}[\rho_{q_+q_0}(0)\\&+\rho_{q_-q_0}(0)]-2e^{-3i\phi(t)\kappa/(4\pi \omega_{\rm q})}[\rho_{q_0q_+}(0)\\&+\rho_{q_0q_-}(0)]\big \}\Big];
    \label{rho00_der}
\end{split}
\end{equation}
\begin{equation}
\begin{split}
    \rho_{11}(t)=&\frac{1}{2}\Big[\rho_{q_+q_+}(0)+\rho_{q_-q_-}(0)
    \\&-e^{-12f(t)\kappa/(4\pi \omega_{\rm q})}\{\rho_{q_+q_-}(0) +\rho_{q_-q_+}(0)\}\Big];
    \label{rho11_der}
\end{split}
\end{equation}
\begin{equation}
\begin{split}
    \rho_{22}(t)=&\frac{1}{3}\Big[1+e^{-12f(t)\kappa/(4\pi \omega_{\rm q})}[\rho_{q_+q_-}(0)-\rho_{q_-q_+}(0)]
    \\&+e^{-3f(t)\kappa/(4\pi \omega_{\rm q})}\big\{e^{3i\phi(t)\kappa/(4\pi \omega_{\rm q})}[\rho_{q_+q_0}(0)\\&+\rho_{q_-q_0}(0)]+e^{-3i\phi(t)\kappa/(4\pi \omega_{\rm q})}[\rho_{q_0q_+}(0)\\&+\rho_{q_0q_-}(0)]\big \}\Big].
    \label{rho22_der}
\end{split}
\end{equation}
If the qutrit 
is initially in the state $\rho_S(0)=|1\rangle\langle1|$, we obtain
\begin{equation}
\begin{split}
    &\rho_{q_+q_+}(0)=\rho_{q_-q_-}(0)=\frac{1}{2};\:\:\:\:
    \rho_{q_-q_+}(0)=\rho_{q_+q_-}(0)=-\frac{1}{2};\\&
     \rho_{q_0q_+}(0) =\rho_{q_+q_0}(0)=0;\:\:\rho_{q_0q_-}(0)=\rho_{q_0q_-}(0)=0;\\&
     \rho_{q_0q_0}(0)=0.
\end{split}
\end{equation}
Using the above  
initial conditions, we obtain the diagonal elements \textcolor{black}{[Eqs.~(\ref{eq:de1})--(\ref{eq:de3})]} as
\begin{equation}
    \langle0|\rho_S(t)|0\rangle=\frac{1}{6}[1-e^{-12f(t)\kappa/(4\pi \omega_{\rm q})}];
\end{equation}
\begin{equation}
    \langle1|\rho_S(t)|1\rangle=\frac{1}{2}[1+e^{-12f(t)\kappa/(4\pi \omega_{\rm q})}];
\end{equation}
and
\begin{equation}
    \langle2|\rho_S(t)|2\rangle=\frac{1}{3}[1-e^{-12f(t)\kappa/(4\pi \omega_{\rm q})}].
\end{equation}
\section{ACKNOWLEDGMENTS}
    
We wish to thank Mikko M\"ott\"onen, Sahar Alipour and Ali Rezakhani for useful discussions. This work has been supported in part by the Academy of Finland Centre of Excellence program Quantum Technology Finland (projects 312298 and 312300).

\end{document}